# Structure and Dynamics of Boron Nitride Nanoscrolls

Eric Perim and Douglas S. Galvao*

Applied Physics Department, Institute of Physics

University of Campinas – UNICAMP, 13083-970, Campinas-SP, Brazil

**Abstract**

Carbon nanoscrolls (CNSs) are structures formed by rolling up graphene layers into a papyruslike shape. CNNs have been experimentally produced by different groups. Boron nitride nanoscrolls (BNNSs) are similar structures using boron nitride instead of graphene layers. In this work we report molecular mechanics and molecular dynamics results for the structural and dynamical aspects of BNNS formation. Similarly to CNS, BNNS formation is dominated by two major energy contributions, the increase in the elastic energy and the energetic gain due to van der Waals interactions of the overlapping surface of the rolled layers. The armchair scrolls are the most stable configuration while zigzag scrolls are metastable structures which can be thermally converted to armchair. Chiral scrolls are unstable and tend to evolve to zigzag or armchair configurations depending on their initial geometries. The possible experimental routes to produce BNNSs are also addressed.

*corresponding author: galvao@ifi.unicamp.br,

Tel. +55-1935215369, FAX: +55-1935215376

## 1. Introduction

Nanometric structures have been intensively studied during the last years, due to new physical phenomena and potential technological applications. Carbon-based materials are among the most investigated in literature. In particular, carbon nanotubes (CNTs)[1] are one of most important subjects in materials science. In spite of more than two decades of intense investigations, the detailed mechanisms of the CNT formation are still an open and polemical question[2]. It has been proposed that CNTs could be a subsequent state of papyruslike carbon structures, generically named carbon nanoscrolls (CNSs) (Fig. 1)[3,4].

Although known since 1960s[3] there are very few works on CNSs. This can be explained in part by the intrinsic experimental difficulties in synthesis and characterization. With the recent advances in low-temperature synthesis[4,5,6] there is a renewed interest in these materials[7,8,9,10,11,12,13]. Like CNTs, CNSs can be made of single or multiple graphene sheets. However, in contrast to CNTs, their diameter can be easily varied (contraction or expansion), making them extremely radially flexible, a very useful property to be exploited to their potential use for Hydrogen storage[14,15,16].

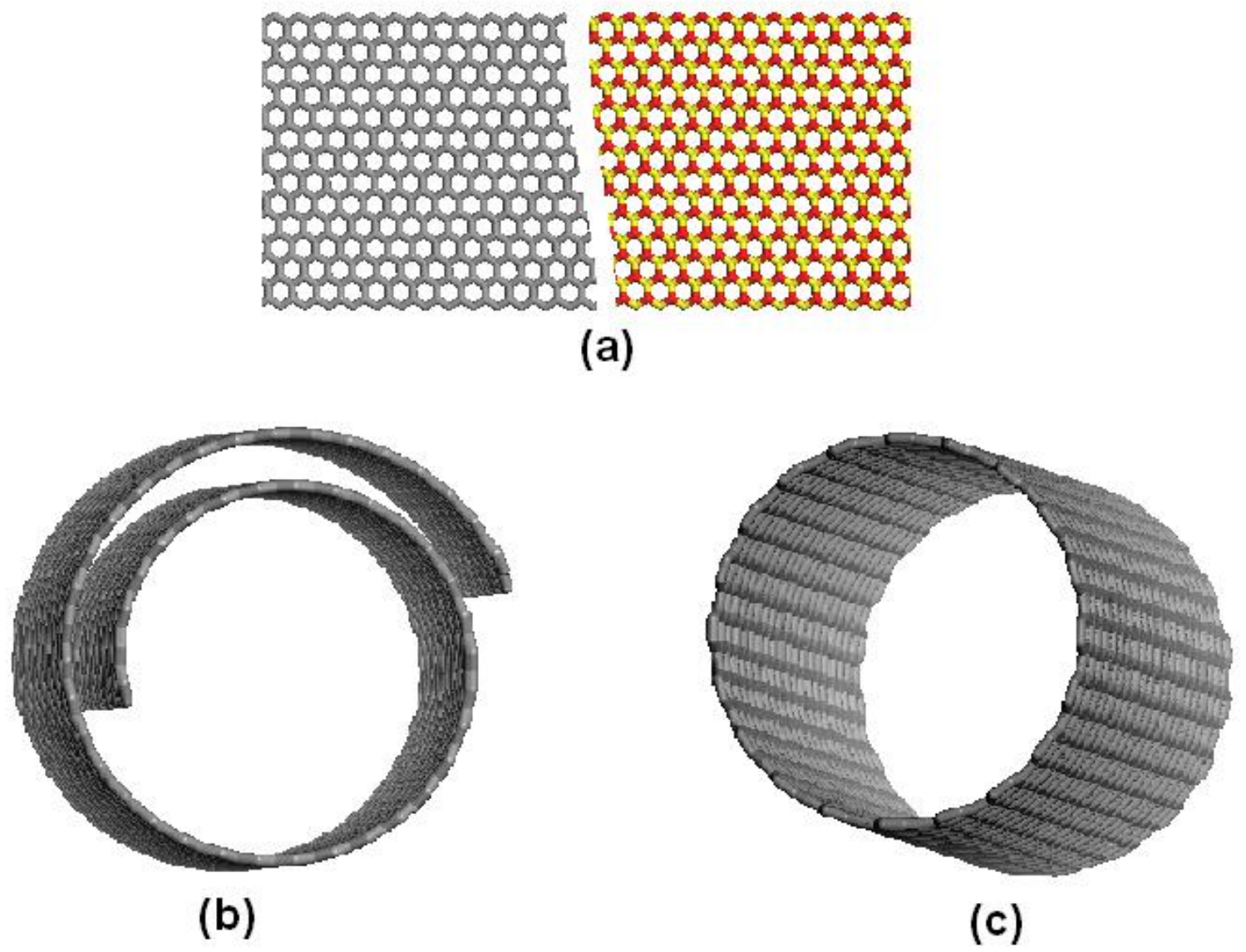

**Figure 1 –** (a) Graphene/BN sheet; (b) Nanoscroll; (c) Nanotube. See text for discussions.

Another important class of nanostructures is boron nitride nanotubes (BNNTs) which were theoretically predicted[17,18] and synthesized with different techniques[19,20]. For a recent review see Goldberg et al [21]. Although BNNTs share some of the remarkable properties exhibited by CNTs, they have some peculiar features such as band gap chirality independence, while CNTs can exhibit semiconducting or metallic characters depending on their helicity[21].

By analogy, CNTs and BNNTs can be formed by rolling up graphite and boron nitride layers, respectively (Fig. 1). For CNSs and boron nitride nanoscrolls (BNSs) a similar analogy may be considered (Fig. 1b).

In principle, the recently reported experimental techniques used to produce CNSs[4,5] can be used to produce BNSs using cubic boron nitride crystals as starting materials. Although a large number of theoretical and experimental works has been reported to BNNTs, BNSs have not been explored so far. In this work we present the first theoretical study of the dynamical and structural properties of BNSs.

Previous molecular mechanics and molecular dynamics studies[8] of different CNSs have shown that their formation results from two competing factors, elastic deformations and van der Waals interactions among overlapping layers. Depending on the geometry and chirality CNSs can be even structurally more stable than their equivalent planar configurations (graphene).

In this work we used the same methodology reported in ref 8 to investigate the dynamical and structural properties of different (size and chirality) BNSs. In particular a critical analysis of similarities and differences among CNSs and BNSs is presented.

## 2. Methodology

In order to address the structural and dynamical aspects of BNSs we have carried out molecular mechanics and molecular dynamics simulations using the well-known molecular universal force field (UFF)[22,23] as implemented in Cerius and Materials Studio[24] software.

UFF is a force field that includes bond stretch, bond-angle bending, inversion, torsions, rotations and van der Waals terms. Electrostatic and solvent effects can be also easily included. UFF has been proved to be very effective in the description of the structural and dynamical aspects of CNTs and CNSs[8,25,26,27].

Boron and nitrogen atoms were assumed as having partial double bonds and $sp^2$ hybridization. No explicit charges were used and the systems are assumed to be neutral. The calculations were carried out with high numerical precision (the used criterion convergence was of $10^{-5}$ kcal/mol in energy and $5x10^{-3}$ kcal.Å/mol for maximum force among atoms). We investigated structures containing up to ~ 20,000 atoms, which precludes the use of full quantum methods. The evolution of the scroll structure was simulated using molecular dynamics (MD) methods for different temperatures (up to 300 K) within the NVT ensemble making use of a Nosé thermostat[28]. In average, each simulation was carried out for a period of 50 ps, with time steps of 1 fs.

The nanoscroll structures were generated by rolling a graphene or BN layer into a truncated Archimedean-type spiral, by rotation around the $y_1$ axis defined by the angle θ shown in Fig. 2. This spiral is defined by the parametric equation

$$r = a\phi + a_0$$

where **a** and $\mathbf{a_0}$ are non-zero constants, and **r** and $\phi$ are the usual cylindrical coordinates. The parameter **a** is related to the initial interlayer spacing **d**, as $\mathbf{a = d / 2\pi}$, where d = 3.4 Å for carbon interlayer distances. Archimedean-type spirals describe well the scroll structures since they have a constant separation distance for successive turnings of the spirals. While multi-layer BNSs are likely to exist, for simplicity we consider here only single layer scrolls.

The angle θ defines the scroll configuration: zigzag for θ=0, armchair for θ=90º, and chiral for $0 < \theta < 90º$. We restricted our analysis of the evolution of the structures during scroll formation for two scroll types: α (Fig. 2b), where there are no uncurled regions of the layer; and β (Fig. 2c), where flat and uncurled regions simultaneously coexist. Although there are many other possible initial structures, the basic structural features can be addressed using only these two structure types as demonstrated for the CNS case[8]. We have considered α and β structures of different sizes (varying H and W values – Fig. 2).

## 3. Results and Discussions

Similarly to CNSs, BNS formation is dominated by two major energetic contributions, the elastic energy increase caused by bending the BN layer (decreasing stability) and the free energy decrease generated by the van der Waals interaction energy of overlapping regions of the layer (increasing stability).

We started analyzing the scroll stability as a function of its internal radius. In Fig. 3 we present the results for a structure of 30 Å x 121 Å that is rolled up with different radius. The change in the configuration energy per atom relative to its equivalent undistorted layer, considered as the reference energy value, ΔE, is also discussed.

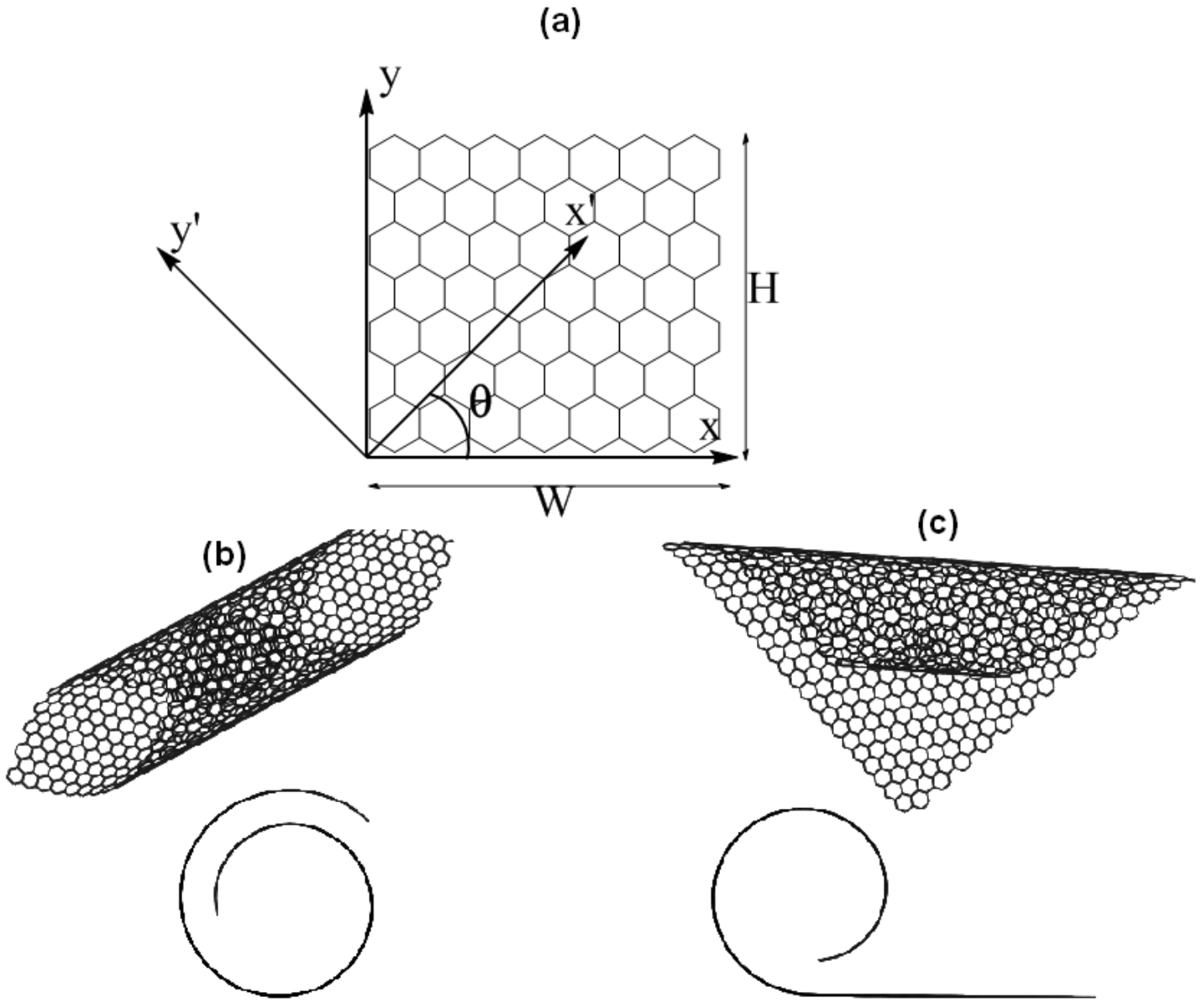

**Figure 2** - (a) The unwrapped honeycomb lattice of a single layer (width $W$ and height $H$). Here $x_1$ and $y_1$ are the scroll axes, which are rotated by an angle θ with respect to the reference coordinate system $xy$. The scroll is generated by wrapping the sheet around the axis $y_1$. Examples of (b) α and (c) β-type CNSs (with θ=45°) and their cross sections are shown. See text for discussions.

As the structure is rolled up and before layer overlap occurs, layer curvature increases and so does the torsion and inversion contributions to the strain energy. Thus the rolled structure becomes less stable in relation to the undistorted (planar) configuration. This implies that the transition from planar to rolled structures must be energy assisted (e. g., through sonication, as experimentally reported for the CNS case[4,5]). If no external energy is provided, the curved structures will return to the more stable planar configurations. When surfaces start to overlap, the van der Waals contributions increase the structural stability.

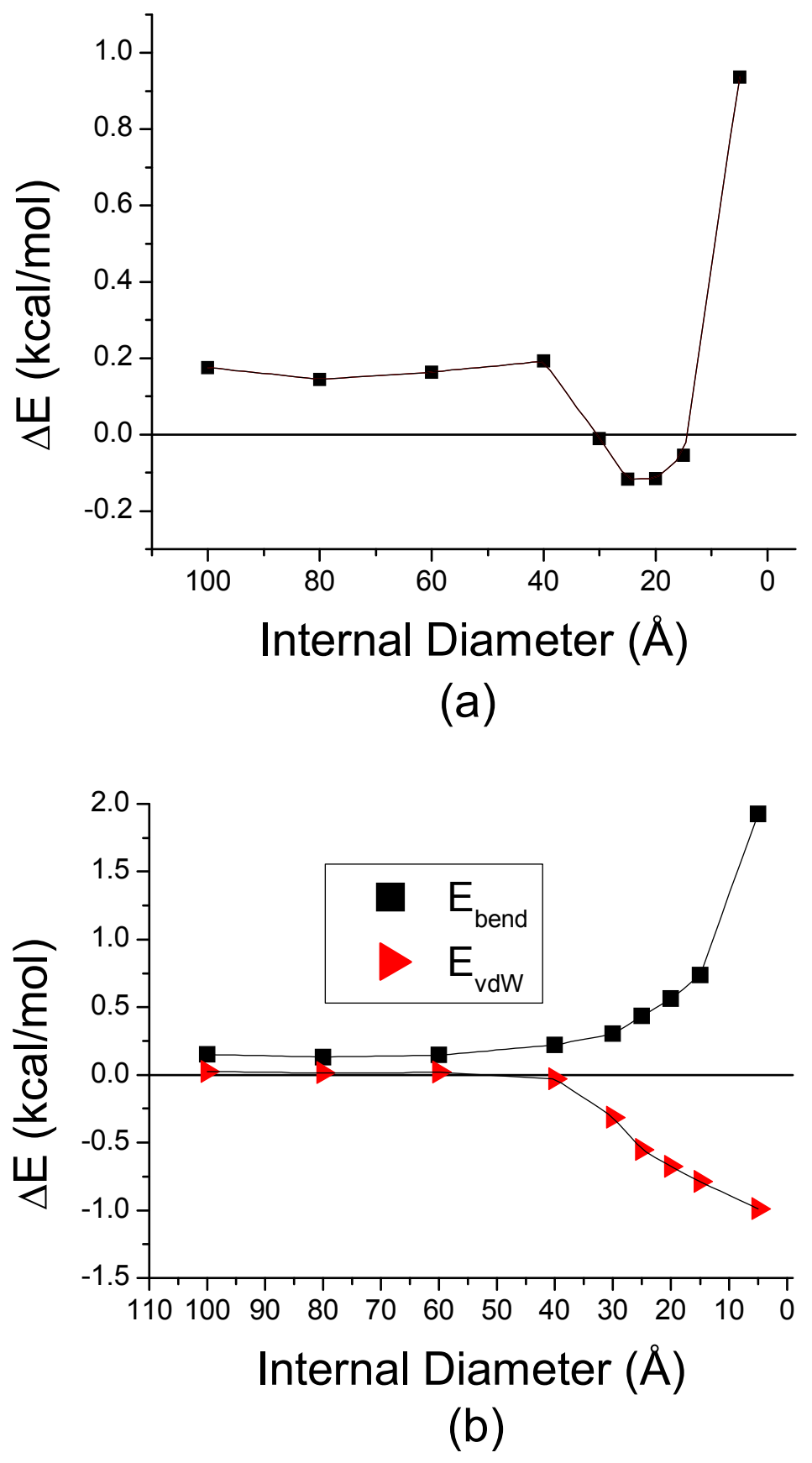

**Figure 3 –** (a) Change in relative total energy (ΔE) during the wrapping of the plane BN sheet into BNNS. (b) contributions of bending and van der Waals terms to the total ΔE.

This process can be better visualized analyzing the van der Waals and bending energy contributions separately (Fig. 3b). For small diameter values the bending contributions outweigh the van der Waals energetic gains and the structure become unstable and tend to evolve to the planar structures. There is a critical minimum inner diameter necessary for the scroll to attain structural stability. Beyond this limit the rolling process would evolve spontaneously and the final structure can be even more stable than its parent planar configuration. For the case displayed in Fig. 3a this occurs for diameter values between 13 and 30 Å. Although these critical values are size and chirality dependent, the general behavior displayed in Fig. 3 is valid for all BNS structures.

We have carried out molecular dynamics simulations in order to address the relative importance of size and chirality in the BNS formation. We considered structures with internal radius above and below the stability line, built from a 100 Å X 100 Å planar structure, in all three configurations (armchair, chiral and zigzag).

In Fig. 4 we present the time evolution energy profile for structures above the stability line which shows that the dynamical behavior is mostly independent of scroll chirality (Fig. 4a). Although oscillations in the scroll structure can exist for some time, they all eventually unfold and converge to planar configurations (see video 4 in the supplementary materials).

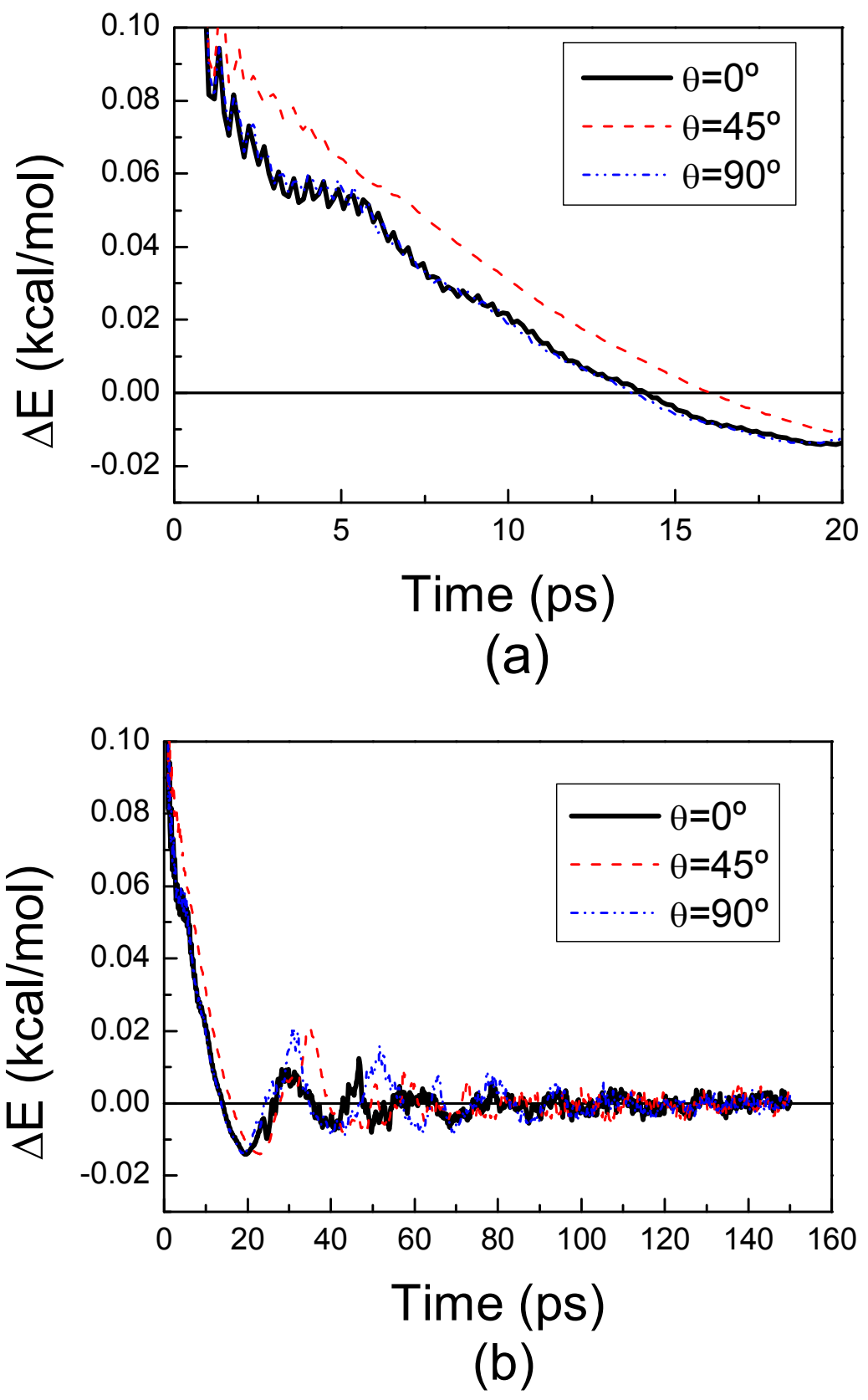

**Figure 4 -** ΔE time evolution profile for scrolls of different chiralities with radius above stability line. (a) Initial temporal stages. (b) Complete temporal process.

In Fig. 5 we present results for structures below the stability line. Although scroll formation is possible for all the different chiralities, their dynamical evolution is differentiated since the onset of the scrolling process (Fig. 5a).

The armchair scrolls are the most stable configurations, even more stable than the planar structures. This aspect can be observed from the first temporal stages (Fig. 5a). Chiral scrolls are the least stable configurations. The simulations show that chiral scrolls can evolve to armchair or zigzag configurations depending on the initial geometries. For the case presented in Fig. 5, the scroll oscillates between zigzag and armchair, but converged to the armchair configuration. Zigzag scrolls, although less stable than the armchair ones, do not easily undergo structural changes (at least for low temperatures) and remain in the zigzag form.

These results indicate that BN armchair scrolls are more likely to be found, especially at high temperatures, where the zigzag scrolls can be thermally converted to armchair ones.

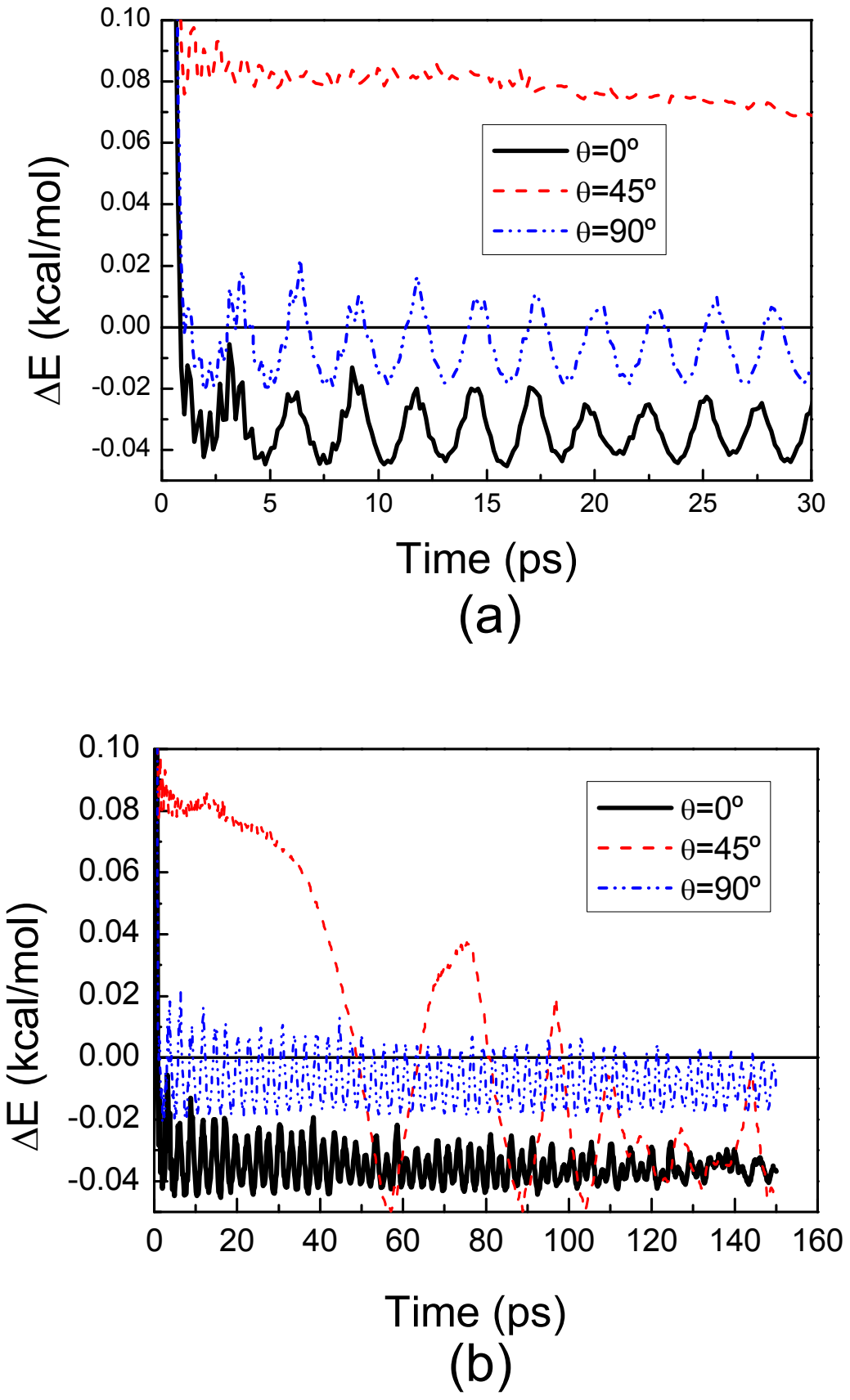

**Figure 5 -** ΔE time evolution profile for scrolls of different chiralities with radius below stability line. (a) Initial temporal stages; (b) Complete temporal processes.

For scrolls formed from rectangular geometries (Fig. 2) BNSs and CNSs present quite similar behavior (Fig. 6) with relation to the stability energies, although for the same dimensions and internal radius BNSs are more stable. This suggests that the van der Waals interactions are more relevant for the BNS than for the CNS case.

The stability energy increases rapidly with the scroll area. This is a direct result from the van der Waals interactions, which are mainly responsible for the stability structural gains and are also proportional to the surface area. For large H and W values the ΔE curve tends towards a constant value.

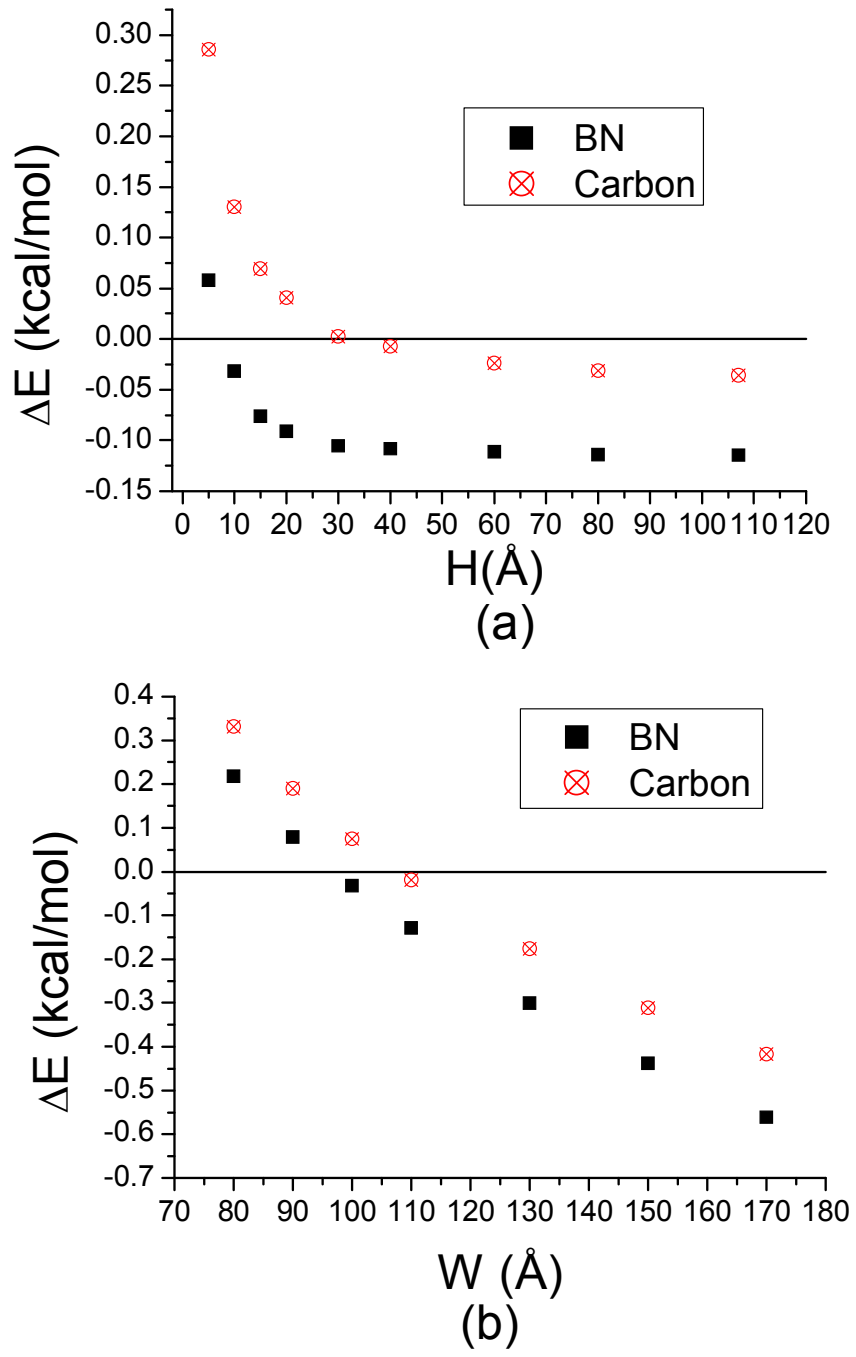

**Figure 6 -** Relative stability for Boron Nitride and Carbon nanoscrolls formed from rectangular sheets with internal diameter ~20Å and (a) W=107Å; (b) H=30Å.

Finally we analyzed the energetics of scroll formation for α- and β-type scrolls from rectangular layers (Fig. 2). Fig. 7 shows the total energy and the van der Waals and bending energy contributions. The α-type configuration is energetically more favorable and the scrolling process occurs faster than for the β-type. This can be seen from the initial energy values and from the slope of the curves (Fig. 7a).

As the α-type allows the simultaneous movement of both overlapped edges in opposite directions, the scrolls are formed faster (see video 3, supplementary materials). This feature can also be seen in Fig. 7b where the fast van der Waals gain for the α-type is clearly observed. The β-type configuration presents a more complex behavior, especially for the bending energy (Fig. 7c), because geometrical changes in the flat layer region occur simultaneously with scroll formation (see video 1 and 5, supplementary materials). It costs less energy to start the bending layer from both extremities (α-type) than to start rolling it from one extremity to the other (β-type). Both types can generate the same BNS by self-rolling processes, but it is more likely (as in the case of sonicated CNSs[4,5]) that the layers can exhibit bending in many directions simultaneously.

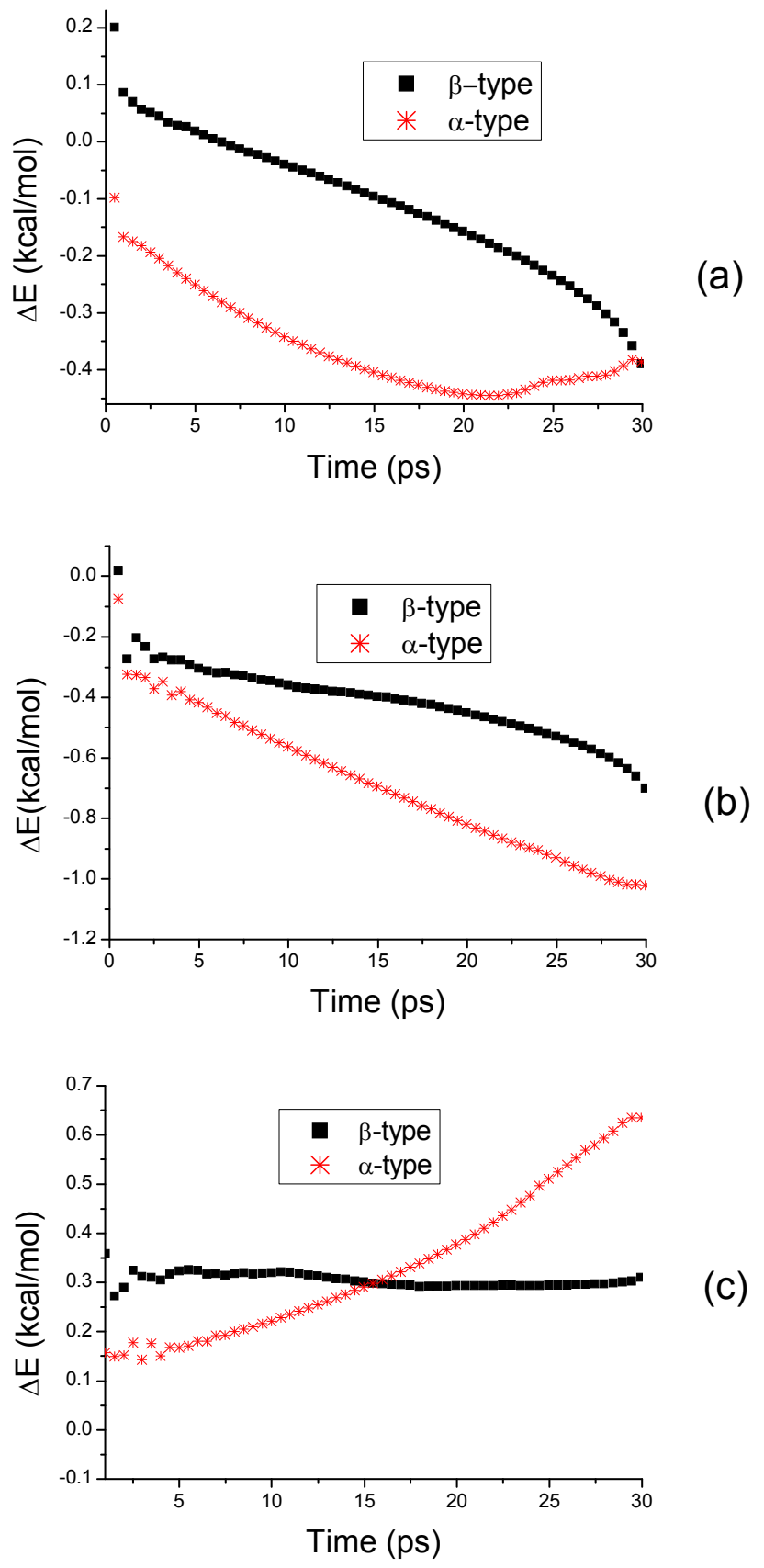

**Figure 7 -** Relative energy time evolution for α and β type scrolls. (a) Total energy; (b) van der Waals energy; (c) Bending energy.

In Figs. 8 and 9 we show conical and non-conical shape BNSs obtained from MD simulations for structures containing ~ 20,000 atoms. We observed that the conical structures are metastable states and their occurrence is size and chirality dependent, with the probability of occurrence being proportional to scroll size. The conical CNSs were experimentally observed[4,5] and our results suggest that they should also occur to BNSs (see video 5, supplementary materials).

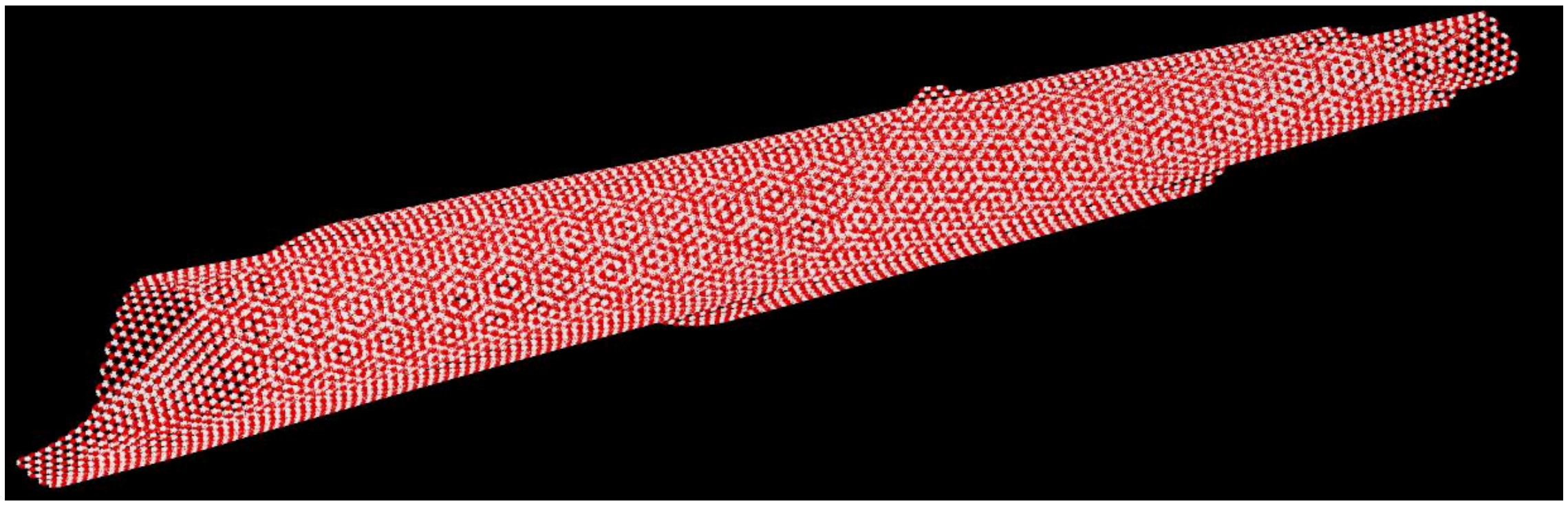

**Figure 8 –** BNNS metastable conical shape.

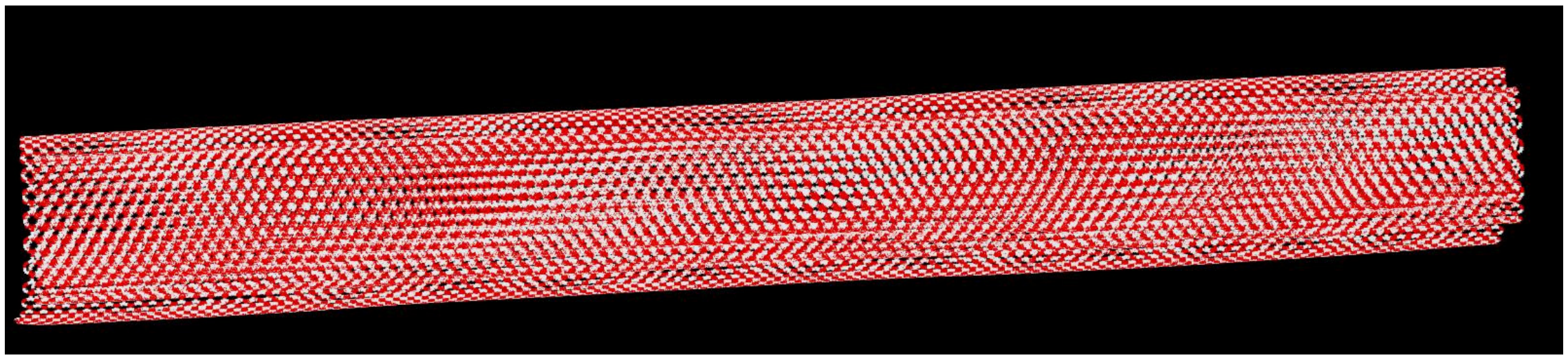

**Figure 9 –** BNNS minimum energy configuration.

## 4. Summary and Conclusions

We have investigated the energetics and dynamical aspects of Boron Nitride Nanoscrolls (BNNSs) using molecular mechanics and molecular dynamics methods. Our results show that the BNNS formation mechanisms are quite similar to those from Carbon Nanoscrolls (CNSs)[8]. Similarly to CNS, BNNS formation is dominated by two major energy contributions, the increase in the elastic energy due to the bending of the initial planar structure (decreasing structural stability) and the energetic gain due to the van der Waals interactions of the overlapping surface of the rolled layers (increasing structural stability). There is a critical diameter for scroll stability; beyond this limit the scrolled structure can be even more stable than its parent planar structures.

The armchair scrolls are the most stable configuration; zigzag scrolls are metastable structures that can be thermally converted to armchair. Chiral scrolls are unstable and tend to evolve to zigzag or armchair scrolls depending on their initial geometries. Similarly to the experimentally observed conical shape CNSs, our results show that BNNS are also likely to be found in these configurations. In principle, the same experimental techniques used to produce CNSs[4,5] could be used to produce BNNSs using cubic boron nitride as starting materials. We believe that our results may stimulate further works along these lines.

**Acknowledgements** – The authors acknowledge financial support from the Brazilian Agencies CNPq, CAPES and FAPESP. We also wish to thank Profs. R. H. Baughman, V. Coluci, S. B. Legoas, F. Sato for helpful discussions, and Prof. M. A. Cotta for critical reading of this manuscript.